# TEM and AFM study of hydrogenated sputtered Si/Ge multilayers

C. Frigeri[1], L. Nasi[1], M. Serényi[2], A. Csik[3], Z. Erdélyi[4], and D. L. Beke[4]

1. CNR-IMEM Institute, Parco Area delle Scienze, 37/A, 43010 Parma, Italy
2. MTA-MFA Institute, Konkoly-Thege ut 29-33, H-1121 Budapest, Hungary
3. Institute of Nuclear Research of HAS, Bem tér 18/C, H-4001 Debrecen, Hungary
4. Department of Solid State Physics, University of Debrecen, P.O. Box 2, H-4010 Debrecen, Hungary

*Corresponding author*: C. Frigeri

CNR-IMEM Institute, Parco Area delle Scienze 37/A, Fontanini, 43010 Parma (Italy), frigeri@imem.cnr.it, fax +39 0521 269206

## ABSTRACT

Multilayers of hydrogenated amorphous ultrathin (3 nm) a-Si and a-Ge layers prepared by sputtering have been studied by atomic force microscopy (AFM) and transmission electron microscopy (TEM) to check the influence of annealing on their structural stability. The annealed multilayers exhibit surface and bulk degradation with formation of bumps and craters whose density and size increase with increasing hydrogen content and/or annealing temperature and time. Bumps are due to the formation of $H_2$ bubbles in the multilayer. The craters are bumps blown up very likely because of too high a gas pressure inside. The release of H from its bonds to Si and Ge occurs within cavities very likely present in the samples. The necessary energy is supplied by the heat treatment and by the recombination of thermally generated carriers. Results by energy filtered TEM on the interdiffusion of Si and Ge upon annealing are also presented.

# 1 - INTRODUCTION

Since a long time amorphous silicon (a-Si), mostly in the hydrogenated state, has been the material of choice for solar cells as well as for other applications in optoelectronics [1]. a-Ge (it, too, mostly in the hydrogenated state) is less employed and studied than a-Si. However, apart from applications as infrared optoelectronic devices and detectors for γ and X rays, it can be used in the solar cell technology, like a-Si, as a low band gap material in tandem solar cells (in the shape of a-Si/a-Ge layer stacks) [2]. Most importantly, it can be alloyed to a-Si to form the a-SiGe alloy whose bandgap depends on the Ge content [3, 4]. This is benificial for tandem and triple solar cells since by using the a-SiGe alloy the band gap structure of the device can be taylored at whish, by changing the Ge composition, for increasing the efficiency by improving the long wavelegth response [3, 5-7]. As a matter of fact, a-SiGe alloys are now extensively used for photovoltaic devices [3, 8]. One way of preparing the hydrogenated a-SiGe alloy is to deposit alternating ultrathin (<5 nm) a-Si:H and a-Ge:H layers. Such multilayer (ML) structure, in fact, can be viewed as a completely mixed hydrogenated a-SiGe alloy [4]. This paper deals with this type of multilayers.

The reason why a-Si and a-Ge as well as a-SiGe alloy are submitted to hydrogenation is that H passivates the dangling bonds of Si and Ge which are well known to introduce deep gap states [9-11]. As a consequence the electrical and optical properties of materials and devices improve significantly [9-11]. For the a-SiGe alloy it was seen that the H content can also change the optical band gap [2, 6]. Despite this the quality of a-Si, a-Ge and a-SiGe alloy is far from being optimized. This is partially ascribed to the behaviour of H itself. For instance, H does not attach to Ge as effectively as it does to Si [3, 4, 9, 12]. A still critical issue of these materials is the light induced degradation due to defect creation (Staebler-Wrosnki effect) [5, 13-15]. Though no conclusive model has been put forward yet to explain such light soaking induced degradation it is generally believed that H could play some important role in it [3, 13-15]. In particular, morphological modifications of the material were reported [13-15]: a surface degradation was observed by Agarwal et al. [13] in light soaked a-SiGe while Han et al. [14] reported on a decrease of the light induced volume expansion with decreasing Si-H concentration in a-Si. The instability of hydrogen with possible formation of internal voids was also suggested even in the absence of light soaking in a-Si by means of effusion experiments [1 and references therein].

Beside light induced degradation, thermally induced degradation is also a key issue of hydrogenated amorphous semiconductors [16] as their electro-optical and structural properties turn out to be unstable against heat treatments that might be necessary, e.g., to dope the semiconductor [16] or to create solid state solutions by interdiffusion in Si/Ge MLs. In the latter case one has also to check how element interdiffusion takes place and what could be the possible influence of the presence of H on it. By using small-angle X-Ray diffraction it was shown by some of the present authors that in amorphous Si/Ge multilayers prepared by DC magnetron sputtering and submitted to heat treatment the diffusion is very asymmetric because of the strong concentration dependence of the interdiffusion coefficient [17-19]. Silicon was seen to enter into the germanium layer but germanium could not diffuse into silicon.

Based on the findings of the references cited in this introduction as well as of others not cited and on our preliminary work structural modifications have also to be expected in Si/Ge amorphous MLs made of ultrathin layers. By using atomic force microscopy (AFM) and transmission electron microscopy (TEM) this work will show that this is really the case. Results obtained by energy filtered TEM (EF-TEM) about the element interdiffusion upon annealing are also presented.

## 2 - EXPERIMENTAL

The multilayers made of alternating ultrathin a-Si and a-Ge layers were prepared by radio frequency sputtering. A conventional apparatus (Leybold Z 400) with basic vacuum pressure of $1 \times 10^{-4}$ Pa was used to deposit them on (100) oriented Si substrates. During sputtering the chamber pressure was kept at 2.0 Pa and the substrate was water cooled. As sputtering gas argon (purity 99.999 %) was used. A 1.5 kV wall potential was applied to the Si and Ge targets. The Si/Ge multilayer stacks consisted of a sequence of 50 couples of Si and Ge layers which were 3 nm thick each as confirmed by TEM. Both reference not-hydrogenated and hydrogenated multilayers were prepared. The hydrogenation of the Si/Ge samples was obtained by adding hydrogen in the sputtering chamber with different flow rates from 0.8 up to 6ml/min. Both not-hydrogenated and hydrogenated multilayers underwent the same heat treatments in high purity argon (99.999%), either of the 1-step annealing type (time 16 h and temperatures of 350, 400 or 430 °C) or of the 2-step type (250 °C for 0.5 h + 450 °C for 5 h).

The structural analyses were carried out by atomic force microscopy (AFM) in the tapping mode and transmission electron microscopy (TEM). Both 2000 FX and 2200 FS Jeol TEMs were employed. The latter one was equipped with an in-column Omega-type Energy Filter and was used to get elemental maps of Si and Ge in the energy filtering operation mode (EF-TEM). Elemental maps where acquired at a magnification of 300k with a GATAN slow scan CCD camera. One post-edge and two pre-edge images were acquired using a slit of 20 eV and 7 eV for Si and Ge, respectively. From the two pre-edge images the background image was evaluated which was then subtracted from the post-edge image. In so doing for each element maps of edge intensity versus the position in the ML stacks were obtained which give a semi-quantitative estimate of its concentration [20]. For TEM only <110> cross-sectional specimens were investigated. They were prepared by mechanical thinning of sandwiches, containing a small piece of the sample, down to 30 µm followed by Ar ion beam thinning at 5-3 keV and 0.35 mA down to electron transparency.

## 3 - RESULTS AND DISCUSSION

No one of the employed microscopies showed any structural anomaly in the not-hydrogenated samples, either in the un-annealed ones or in the annealed ones. The surface (checked by AFM) and the bulk (checked by TEM) of the samples were of pretty high quality even after annealing in the most severe conditions used in this work as can be seen in the TEM and AFM images of Fig. 1 and 2, respectively. In particular, in all the annealed not-hydrogenated samples the rms value at the surface was extremely low (~0.19 nm) and practically the same as for the un-annealed samples.

Morphological and structural changes were instead detected for all the hydrogenated samples submitted to annealing. The AFM images of Fig. 3 give examples of the status of the sample surface after annealing. The surface flatness degrades in an appreciable way by formation of bumps and craters whose density and dimension increase with hydrogen content and, for the same hydrogen content, with increasing annealing temperature and/or time. In the sample with the smallest H content undergone to the lightest annealing treatment (350 °C, 16h) the bumps and craters cover ~4% and ~$1^{x}10^{-2}$ % of the sample surface, respectively. On the other hand, such percentages of degraded areas increase with increasing H content and annealing temperature reaching values as high as ~36% and ~19% for bumps and craters, respectively, in samples like the one of Fig. 3 b) of 6 ml/min hydrogen flow rate and annealed at 250 °C, 0.5h + 450 °C, 5h. Fig. 3 b) suggests that the craters correspond to areas where the surface has blown up with destruction of the material. Since the craters are empty it can be hypothesised that they contained some gas, namely hydrogen, and that they exploded up when the inner pressure had reached some value. Under such an assumption the bumps correspond to zones where the hydrogen has accumulated during the annealing treatment which have not yet blown up, i.e., they formed earlier that the craters. The structural changes in the bulk of the Si/Ge ML stack occurring because of annealing in our hydrogenated samples are visible in the TEM image of Fig. 4 which shows a ML area where the layers are no longer distinguishable. The arrow indicates the place where it reaches the top surface. Outside this area the layer sequence is regular. Such zones where the layer structure has disappeared should be the first stage of hydrogen bubble formation.

The formation of the hydrogen bubbles should initiate at very small cavities of nanometer size very likely present in the amorphous Si and Ge layers. According to the findings reported by Beyer [1 and references therein] about investigations of the effusion of hydrogen from a-Si, hydrogen desorbs from the internal surfaces of the small cavities by rupture of two Si-H bonds with formation of a $H_2$ molecule, as displayed in Fig. 5. Likewise, the same occurs for the Ge-H bond. Alternative and complementary to this mechanism the formation of the $H_2$ molecule can also take place by rupture of just one Si-H (Ge-H) bond the second H atom being supplied by the population of the H interstitials certainly present in the Si/Ge ML stacks, especially in the Ge layers since H bonds to Ge less efficiently than to Si [3, 4, 9, 12].

The energy necessary to break the H bonds to Si and Ge should be supplied by the thermal treatment both directly, as just thermal energy, or indirectly through the thermally generated carriers [16]. In a-Si [16] and a-SiGe [13] the distribution of H is not uniform which causes local band gap variations, so-called elastic fluctuations of the band gap, the zones of less H content having the smaller band gap [13, 16]. Voids where H has been released from Si (or Ge) atoms can be in our samples preferential sites where band gap fluctuations occur. Such non-uniform band gap may cause the thermally generated electron hole pairs to drift to the small band gap regions and recombine there [16]. The energy released in the recombination process can thus be used to break additional Si-H and Ge-H bonds. Energy gain from recombination of photogenerated electron hole pairs as suggested in the case of light soaking [13] is expected to play a quite minor role as our annealings were performed under laboratory room illumination and negligible photogenerated carriers should have been created. The release of a H atom from the host lattice atom is expected to be more likely in the Ge layers since the bond energy of Ge-H is lower than the one of Si-H, namely 69 kcal/mole (2.99 eV) vs 76 kcal/mole (3.29 eV) [21]. This hypothesis can be supported by the work of Xu according to which the Ge-H bond evolution occurs at lower temperature than for the Si-H bond, the Ge-H bond being unstable just above 175 °C [12].

Interconnections between the cavities will favour the diffusion of the $H_2$ molecules [1] and their gathering together with associated formation of larger and larger bubbles. Because of the higher volume of the $H_2$ molecule with respect to the one occupied by two hydrogen atoms bonded to Si or Ge, the host Si and Ge lattices will be strained with possible macroscopic deformation of the Si/Ge ML stack depending on the H content. At present we do not have experimental evidence for this. However, the main contribution to the growth of the $H_2$ bubbles up to the formation of micron size bumps and eventually craters comes from the temperature, for the same hydrogen content. The contribution of the annealing temperature is three-fold. Firstly, it supplies the energy to break the Si-H and Ge-H bonds, as said above. Secondly, it enhances the diffusion of the $H_2$ molecules. Last but not least, it makes the volume of the $H_2$ gas bubbles to increase according to the gas law. This could explain the results shown above, i.e., formation of mostly only small surface bumps (internal bubbles) for small annealing temperatures and the remarkable high density of craters (exploded bubbles/bumps) for high annealing temperatures.

The influence of annealing on the intermixing of Si and Ge between the layers was checked as well by using EF-TEM. Results are given in Fig. 6 for a hydrogenated sample (flow rate 3 ml/min) annealed at 350 °C. It can be seen that Ge is strongly localized at given layer positions, i.e., its own regular locations as if it had not been annealed. On the other hand, the boundaries of the Si layers in the EF-TEM map are less sharp indicating that Si could have diffused out of its original place into the Ge layer after annealing. This confirms the results obtained in previous work about the asymmetry of the diffusion process [17-19]. This also shows that element interdiffusion occurs despite the partial destruction of the ML structure because of the formation of hydrogen bubbles and related bumps and craters. It is still not clear whether and in which way hydrogen can influence diffusion (accelerates it or slows it down compared to the not hydrogenated Si/Ge multilayers). Because H has a lower binding energy to Ge, we suppose that during annealing the H starts to be released firstly from this layer. This yields the formation of more dangling bonds in the Ge layers and the mixing of layers could be speeded up. Further work is necessary to clarify this point.

## 4 - CONCLUSIONS

Multilayers of hydrogenated amorphous ultrathin (3 nm) a-Si and a-Ge layers prepared by radio frequency sputtering are structurally unstable against heat treatments as hydrogen forms bubbles inside the ML which give rise to bumps and craters for high H content and/or temperature. In the worst case here considered the bumps and craters cover up to 36% and 19%, respectively, of the sample surface. The H starts to be released at the surfaces of small internal cavities. Very likely this first occurs in the Ge layers because H has a lower binding energy to Ge than to Si. Every $H_2$ molecule included in the bubbles is due to the rupture of two Si-H (or Ge-H) bonds. Alternatively, rupture of one bond only takes place the second H atom being supplied by a H atom in interstitial position. Such thermally activated rupture of the Si-H and Ge-H bonds depassivates the dangling bonds as well. Worsening of the electro-optical properties of the MLs has thus to be expected. Annealing of the MLs also causes element interdiffusion that turns out to be asymmetric.

**ACKNOWLEDGEMENTS**

This work was supported by the Scientific Cooperation Agreement between MTA (Hungary) and CNR (Italy).

**FIGURE CAPTIONS**

Fig. 1 – Typical TEM image of not-hydrogenated samples submitted to annealing. Similar images were obtained for the un-annealed samples as well. Arrow points to the top surface.

Fig. 2 - Typical AFM image of annealed not-hydrogenated samples. rms = 0.19 nm on the whole sampled area. Similar images were obtained for the un-annealed samples as well.

Fig. 3 - AFM images of the surface of annealed hydrogenated samples: a) hydrogen flow rate 0.8 ml/min, 1-step annealing at 350 °C: small bumps are visible. b) hydrogen flow rate 6 ml/min, 2-step annealing: big bumps and craters are visible.

Fig. 4 - TEM image of an annealed hydrogenated sample (hydrogen flow rate 3 ml/min, 1-step annealing at 350 °C). An area is visible in the centre of the ML where the layer structure was destroyed. White arrow indicates where it touches the top surface. Regular layer sequence is visible at top, bottom and left sides of that area. The ML is here detached from the Si substrate.

Fig. 5 – Sketch displaying the rupture of two Si-H bonds leading to the formation of a $H_2$ molecule. The same holds for the Ge-H bond.

Fig. 6 – Hydrogenated sample annealed at 350 °C, hydrogen flow rate 3 ml/min. a) TEM image. b) and c) corresponding EF-TEM map for Si and Ge, respectively. The area shown here is the top of the ML stack. The bottom part of the pictures is the glue used to prepare the cross-section specimen, i.e., the sample is here shown upside down with respect to the other TEM pictures of the paper.

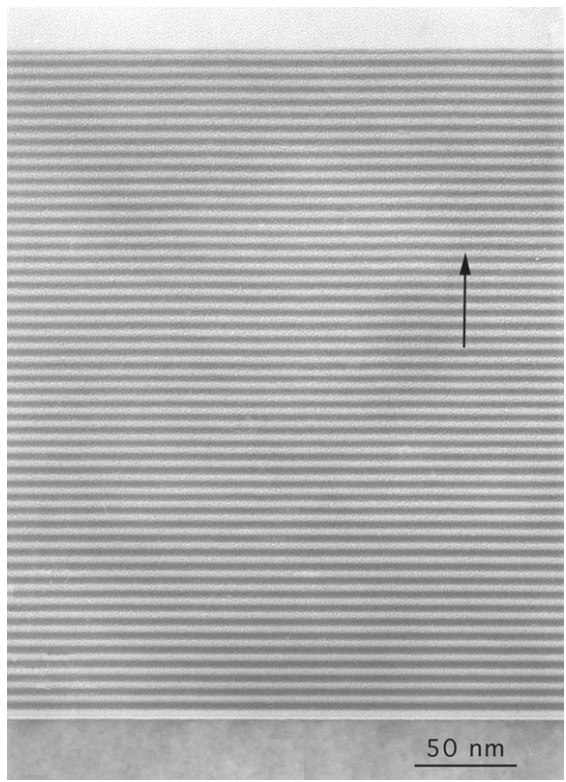

**Fig. 1**

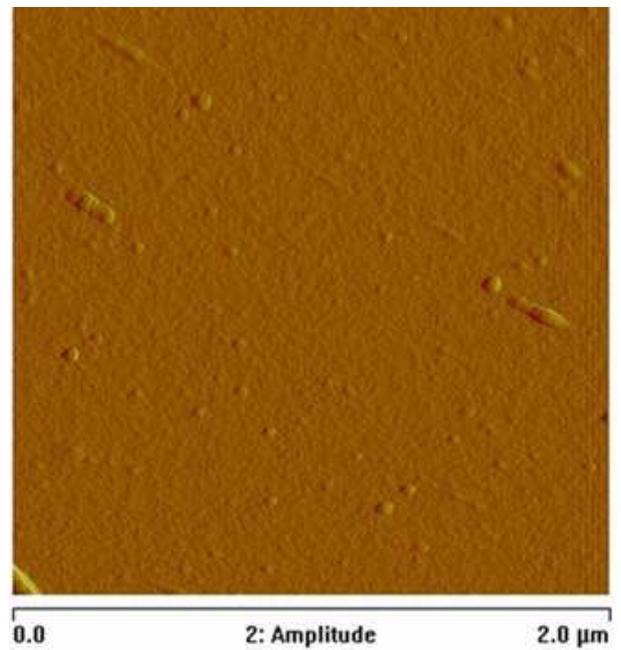

**Fig. 2**

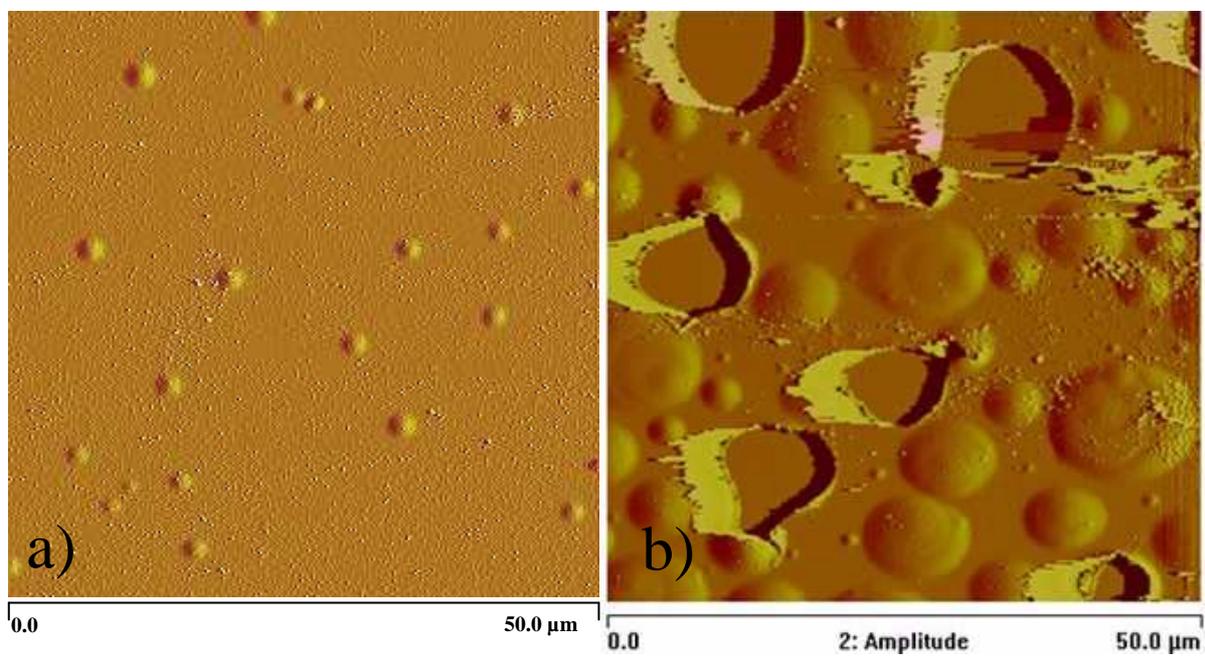

**Fig. 3 a) - b)**

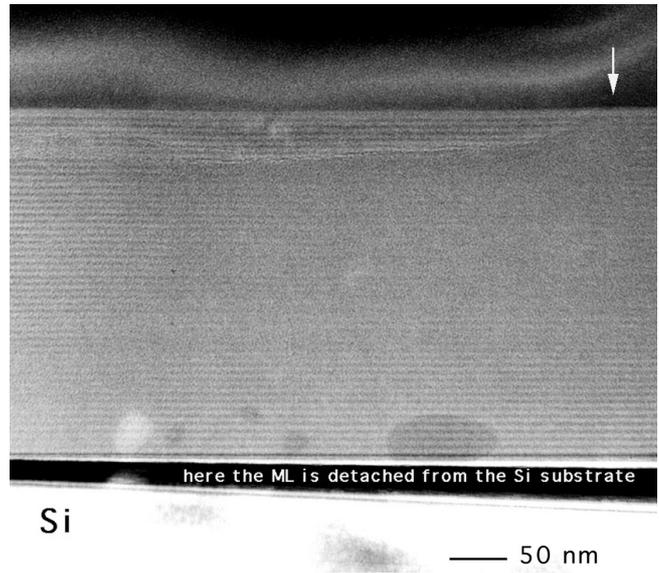

**Fig. 4**

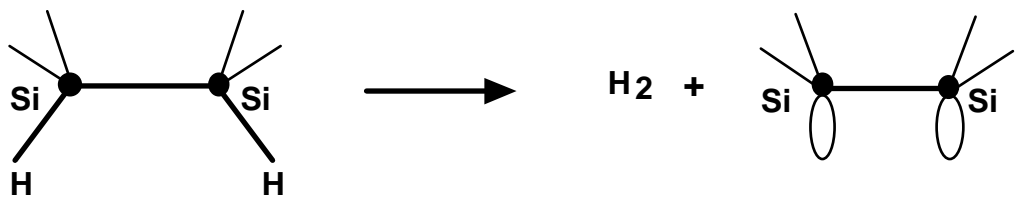

**Fig. 5**

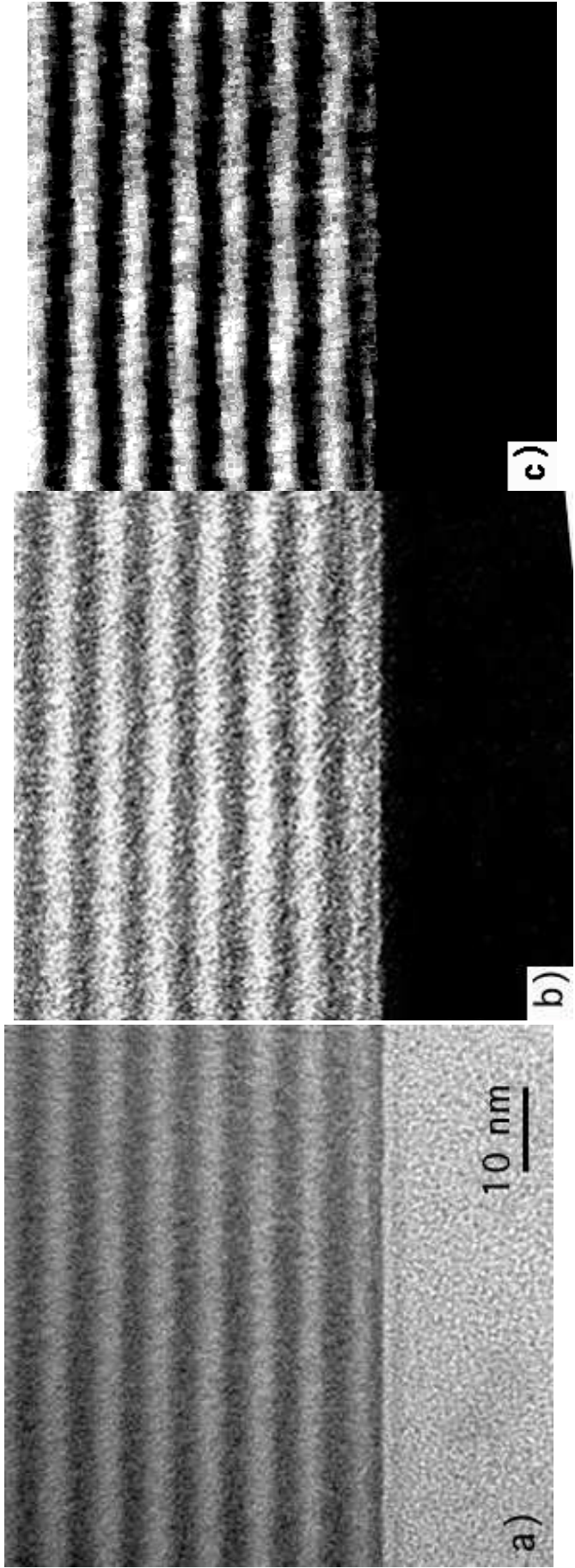

**Fig. 6**